# Velocity-dependent optical forces and Maxwell's demon

J.D. Franson

University of Maryland Baltimore County, Baltimore, MD 21250 USA

An atom placed in a focused laser beam will experience a dipole force due to the gradient in the interaction energy, which is analogous to the well-known optical tweezers effect. This force will be dependent on the velocity of the atom due to the Doppler effect, which could potentially be used to implement a Maxwell's demon. Photon scattering and other forms of dissipation can be negligibly small, which would seem to contradict quantum information proofs that a Maxwell's demon must dissipate a minimum amount of energy. We show that the velocity dependence of the dipole force is cancelled out by another force that is related to the gradient in the phase of the laser beam. As a result, a Maxwell's demon cannot be implemented in this way.

## I. INTRODUCTION

Maxwell's hypothetical demon has spurred the imagination of physicists for 150 years [1-3]. There has been considerable interest in classical models [3-5] for Maxwell's demons as well as those operating at the quantum-mechanical level [6-9]. In all cases, the operation of a Maxwell's demon must dissipate a minimum amount of energy on the order of $kT$, where $k$ is Boltzmann's constant and $T$ is the temperature [6, 10-11]. In this paper, we investigate the possibility of implementing a Maxwell's demon using the dipole force [12-20] exerted on an atom in a focused laser beam, where the energy dissipation can be negligibly small.

The energy of an atom in a laser beam will be shifted by an amount $U(\mathbf{R})$ due to the interaction of the dipole moment of the atom with the electric field of the laser, where $\mathbf{R}$ is the location of the atom. In a focused laser beam, the gradient in the intensity of the field produces a corresponding gradient in the effective potential $U(\mathbf{R})$. This results in a dipole force [13, 14] on an atom that is analogous to the well-known optical tweezers effect [21]. If an atom is moving along the direction of the laser beam, the Doppler effect will produce a velocity dependence of the dipole force that could potentially be used to implement a Maxwell's demon as described in more detail in the next section.

Photon scattering and other forms of energy loss can be negligible, and the dipole force is generally considered to be a coherent process with no inherent dissipation as a result. This raises the question of whether or not the dipole force could be used to implement a Maxwell's demon without dissipating a significant amount of energy. We resolve this "paradox" by showing that there is another velocity-dependent force on an atom that is related to the gradient in the phase of the laser beam rather than the gradient in the intensity. The velocity dependence of these two forces cancel out in such a way that a Maxwell's demon cannot be implemented. This situation provides additional insight into the fact that a minimum energy on the order of $kT$ must be dissipated in the operation of a Maxwell's demon.

We begin with a brief review of Maxwell's demon and the dipole force exerted on an atom in a focused laser beam. The classical motion of an atom is then calculated using Newton's laws combined with the usual expression for the dipole force. These classical calculations show that the dipole force on an atom is velocity-dependent and could potentially be used to implement a Maxwell's demon. Schrodinger's equation is then solved numerically, and the results show that quantum mechanics does not predict a velocity-dependent force on an atom along the direction of propagation of a focused laser beam. The quantum-mechanical motion of the atom is then calculated analytically in the Heisenberg picture, where it can be seen that the dipole force due to the gradient in the intensity of the laser beam is cancelled by another force associated with the gradient in the phase of the field. The combined effect of these two forces does not allow the implementation of a Maxwell's demon in this way, which is consistent with the fact that the dipole force need not dissipate a significant amount of energy.

## II. MAXWELL'S DEMON

A variation on Maxwell's original demon [1-2] is illustrated in Fig. 1, where two chambers initially contain an equal number of gas atoms. All of the atoms are moving with thermal velocities on the order of 1 km/s as described by the Maxwell-Boltzmann distribution at room temperature. A miniature trap door controlled by the demon is opened when an atom arrives from the chamber on the right, allowing it to be transferred into the chamber on the left. But the demon closes the trap door when an atom arrives from the chamber on the left, thus preventing its transfer into the other chamber. All of the atoms will eventually end up in the chamber on the left, in which case the entropy of the system will be reduced. Maxwell's original demon was assumed to allow higher-energy atoms to accumulate on one side of the partition, which is similar in nature to the situation considered here.

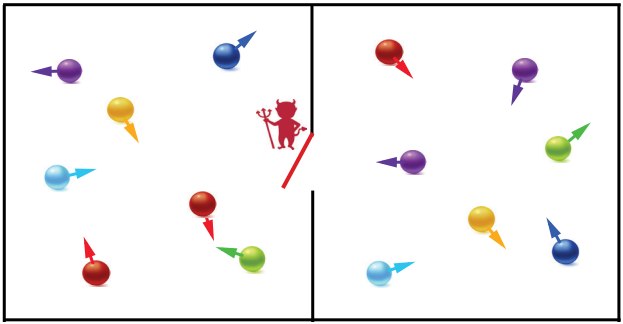

**FIG. 1. Maxwell's demon.** A hypothetical "demon" controls a miniature trap door. The demon opens the trap door when an atom approaches from the chamber on the right, allowing it to pass into the chamber on the left. The demon closes the trap door when an atom from the chamber on the left attempts to pass into the other chamber. All of the atoms will eventually end up in the chamber on the left, which would reduce the entropy of the system and could be used to produce useful work. A potential implementation of a Maxwell's demon using a focused laser beam is shown in Fig. 2 below.

Useful energy could be extracted by using the partition between the two chambers as a piston, for example [3]. Arguments based on classical [10, 11] and quantum [6-8] information theory show that the operation of a Maxwell's demon requires an energy cost of approximately $k_B T/2$ per operation, where $k_B$ is Boltzmann's constant and $T$ is the temperature. This includes the resources required to measure the state of the system and erase the memory used by the demon. In that case, the total entropy would not be reduced and the second law of thermodynamics would be upheld.

An example of a potential implementation of a Maxwell's demon is shown in Fig. 2. As discussed in more detail below, a focused laser beam can exert a repulsive dipole force on an atom that is dependent on its velocity as a result of the Doppler shift. An atom travelling to the left along the center of the laser beam will experience a relatively small force and is more likely to have enough energy to pass through the high-intensity region of the field near the focal point. An atom travelling to the right, however, will experience a larger force and is more likely to have its velocity reversed and end up on the same side where it originated. The effect on the atoms is similar to that of a Maxwell's demon, although the process is not 100% efficient and atoms with a very high energy can pass through the focus of the laser beam in either direction, for example. Nevertheless, such a process would concentrate the number of atoms on one side of the apparatus and reduce the entropy of the system.

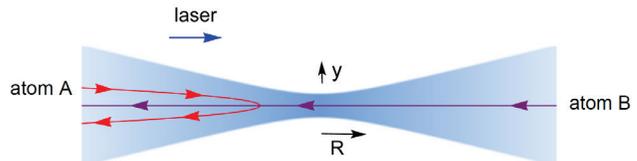

**FIG. 2. Maxwell's demon using a focused laser beam and the dipole force.** For blue detunings, the interaction between the dipole moment of an atom and the electric field of a laser beam creates a repulsive potential that is largest at the point of highest intensity. An atom initially moving in the same direction as the laser beam (red curve) will experience a relatively large repulsive force due to the Doppler shift and its velocity will be reversed. An atom moving in the opposite direction (purple curve) with the same initial energy will experience a smaller repulsive force and it will be able to pass through the high-intensity region at the focus and on to the other side. This velocity dependence could be used to implement a Maxwell's demon if the laser is focused through the center of an aperture between two chambers as in Fig. 1. Photon scattering and other forms of energy dissipation can be made negligibly small for the dipole force.

A Maxwell's demon has already been implemented using another optical approach that involves photon scattering [9]. What is intriguing about the example in Fig. 2 is the fact that photon scattering and other forms of energy dissipation can be made negligibly small by detuning the frequency of the laser beam or by choosing atoms with a long radiative decay time as discussed below. This example is intended to serve as a "paradox" of sorts: How would it be possible for a Maxwell's demon of this kind to operate with negligible dissipation, given the quantum-information proofs to the contrary? The answer to this question is discussed in Section VI, where it is shown that the velocity dependence of the dipole force is cancelled out by another force that depends on the gradient in the phase of the laser.

### III.  DIPOLE FORCES

The dipole force of interest here is proportional to the gradient in the intensity of the light, as illustrated in Fig. 2 [13, 14]. In addition to the dipole force, the incoherent scattering of photons can also exert a force on an atom as illustrated in Fig. 3 [22-24]. The dipole force need not produce any significant dissipation while the effects of photon scattering can be made negligibly small in the limit of large detunings, as described below.

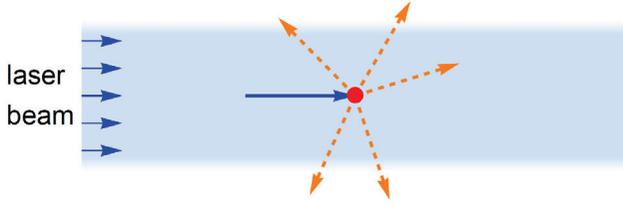

**FIG. 3. The scattering force exerted on an atom in a uniform (unfocused) laser beam.** Here the dipole force of Fig. 2 does not contribute and the only force on an atom is due to the scattering of photons in the original laser beam (blue arrow) into other directions (orange arrows). This effect is commonly used to cool atoms [22-24], but energy is dissipated in the process unlike the dipole force of Fig. 2 where photon scattering can be negligibly small.

The dipole forces of interest here are due to the interaction of the electric field $\mathbf{E}(R)$ of the laser beam with the dipole moment $\mathbf{d}$ of the atoms. For simplicity, we will assume that the atoms are confined to linear motion along the central axis of the laser beam, with $R$ the distance from the center of the focal point as illustrated in Fig. 2. The interaction Hamiltonian is then given as usual by $\hat{H}' = -\hat{\mathbf{d}} \cdot \mathbf{E}(R) = -qyE_y(R)$, where $q$ is the charge of an electron [25]. We have chosen the polarization of the electric field to lie along the y axis of Fig. 2 and we will assume that the strong laser field can be treated classically. If the interaction is sufficiently weak that we can use perturbation theory, then the energy of the atoms will be shifted by an amount $U(R) = |M|^2 / \hbar \Delta$, where $M$ is the matrix element of $\hat{H}'$ between the two atomic states and $\hbar$ is Planck's constant divided by $2\pi$ [25]. The detuning $\Delta$ for an atom at rest is defined by $\Delta = (\omega_L - \omega_A)$, where $\omega_L$ is the angular frequency of the laser beam and $\omega_A$ is the angular frequency of the transition between the two relevant atomic energy levels. $U(R)$ serves as an effective potential energy for the atoms and the dipole force $f(R)$ on the atoms is equal to $-\nabla U(R)$ in the limit of weak fields. Thus the dipole force is proportional to $1/\Delta$ and it is attractive for $\Delta < 0$ (red detuning), while we will be interested in the repulsive potential that occurs for $\Delta > 0$ (blue detuning). The attractive potential for red detunings is analogous to the optical tweezers technique that is widely used to manipulate dielectric particles [21].

The angular frequency $\omega_L'$ of the laser light as seen by a moving atom is given by $\omega_L' = \omega_L + \delta\omega_D$, where $\delta\omega_D = -2\pi v / \lambda$ is the Doppler shift, $v$ is the velocity of the atom in the direction of the laser beam, and $\lambda$ is the wavelength of the light. (We ignore relativistic terms on the order of $(v/c)^2$ throughout, where $c$ is the speed of light.) Thus the dipole force on a moving atom will be velocity dependent and inversely proportional to $\Delta' = (\omega_L + \delta\omega_D - \omega_A)$ instead of $1/\Delta$. This allows an atom incident from the right to pass through the laser focus while an atom incident from the left with the same initial energy will be reflected, as illustrated in Fig. 2. This situation only occurs over a limited range of initial velocities, but it would allow a focused laser beam to function as a Maxwell's demon that allows more atoms to pass in one direction than the other.

For comparison, the rate of photon scattering is given by $P_E / \tau$, where the probability $P_E$ that an atom is in its excited state is $P_E = |M / \hbar\Delta'|^2$ in the perturbative limit [25] and $\tau$ is the radiative lifetime of the atom. The scattering rate can be made negligibly small by choosing a large detuning as is widely done in practical applications, since the dipole force is proportional to $1/\Delta'$ while the scattering rate is proportional $1/\Delta'^2$. As a result, the dipole force is generally considered to be a coherent process with no intrinsic dissipation. Alternatively, the scattering rate can be reduced to a negligible level by choosing atomic states with a sufficiently long lifetime; photon scattering will be negligible if $\tau$ is much larger than the transit time of an atom through the laser beam.

We will consider the limit of large $\tau$ where photon scattering is negligible. (Other authors [19] have considered the opposite limit.) We will also assume that the focal length of the laser beam (Rayleigh length) is sufficiently large that the Hamiltonian is slowly varying and the adiabatic theorem applies, in which case there is no dissipation due to a permanent change in the atomic state. No significant amount of energy is dissipated under these conditions.

An unfocused laser beam can also exert a force on an atom as illustrated in Fig. 3, but in that case the force is proportional to $1/\Delta'^2$ and it is due entirely to the scattering of photons [22-24]. A Maxwell's demon has previously been demonstrated using the scattering of photons, where it was shown that an energy $\sim k_B T/2$ per operation must be dissipated in that case [9]. The use of a focused laser beam and the dipole force would potentially allow a similar operation but with negligible dissipation.

The dipole force along the direction of the laser beam depends on the velocities of the atoms in that direction. Magnetic forces are also velocity dependent, but there the force in any given direction depends only on the velocity in orthogonal directions. As a result, magnetic forces satisfy Liouville's theorem which precludes the operation of a Maxwell's demon in thermal equilibrium, whereas the dipole forces of interest here do not.

## IV. CLASSICAL CALCULATIONS

Figures 4(a) through 4(c) show the results of a classical calculation in which Newton's laws were combined with the dipole force, which was assumed to be given by $f_R(R) = -\nabla U(R)$. The details of the numerical calculations are described in the Appendix.

For simplicity, the atom was assumed to travel along the center of the laser beam with an initial velocity of 2 km/sec at a distance of $R = -200 \,\mu\text{m}$ from the focal point of the laser, which had a Rayleigh length of $100 \,\mu\text{m}$. The mass of the atom was arbitrarily taken to be $3.3 \times 10^{-31}$ kg in order to facilitate the numerical quantum-mechanical calculations described below. The laser detuning was $\Delta = 2\pi \times 5$ GHz and the matrix element $M$ of the interaction Hamiltonian was chosen to be $\hbar \Delta / 5$. The wavelength of the laser was $\lambda = 2 \,\mu\text{m}$, which gives a Doppler shift of $2\pi$ GHz at the initial velocity of 2 km/s. The Doppler shift corresponds to 20% of the detuning and we would therefore expect a comparable velocity dependence of the force initially, with a smaller Doppler shift as the atom is slowed down by the laser beam.

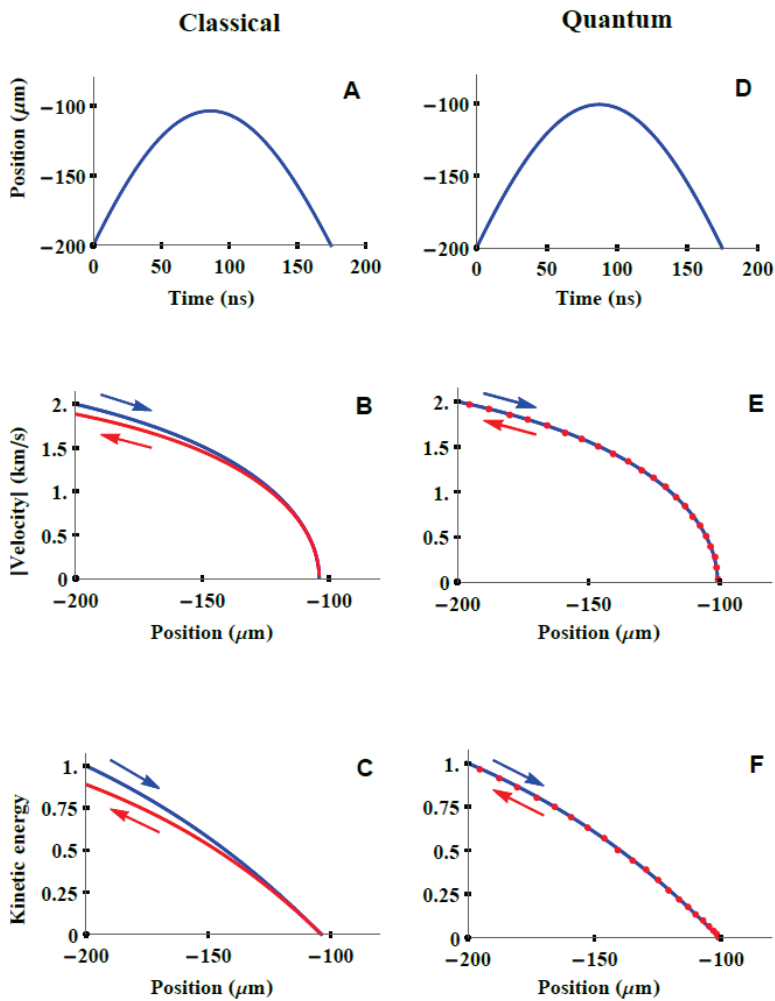

FIG. 4. **Trajectory of an atom moving in a focused laser beam.** The classical trajectory calculated using Newton's laws of motion and the dipole force is shown in (a) through (c), while the corresponding results calculated numerically from Schrodinger's equation are shown in (d) through (f). The calculated position as a function of time is shown in (a) and (d). The absolute value of the velocity as a function of position is shown in (b) and (e), where the blue (upper) curve corresponds to the motion towards the focal point and the red (lower) curve corresponds to the motion away from the focal point after the velocity has been reversed. The kinetic energy (normalized to its initial value) is shown in (c) and (f). It can be seen that the final kinetic energy is reduced in the classical calculations but not in the quantum-mechanical case, which shows that quantum mechanics does not give a velocity-dependent optical force.

Fig. 4(a) shows the classical position of the atom as a function of time. It can be seen that the repulsive dipole potential is sufficient to reverse the atom's velocity for this set of initial parameters. The absolute value of the velocity is plotted in Fig. 4(b) as a function of the position, where the blue (upper) curve corresponds to the trajectory moving in towards the focal point while the red (lower) curve corresponds to the trajectory moving away from the focus. It can be seen that the magnitude of the velocity is less at the end of the process than it was at the beginning, which is due to the fact that $U(R)$ is not a conservative potential. This is even more visible in the plot of the kinetic energy normalized to its initial value in Fig. 4(c). This change in the kinetic energy is a hallmark of a velocity-dependent force.

The classical change in the kinetic energy of an atom due to the dipole force is plotted as a function of its initial energy in Fig. 5. Both the initial and final kinetic energies were calculated at a large distance from the focal point where $U(R)$ was negligible. An atom with a sufficiently large initial energy to pass through the focal point will experience no net change in kinetic energy due to the symmetry of $U(R)$. It can be seen that more atoms incident from the left than the right will have their velocity reversed and undergo a net change in kinetic energy, as was illustrated schematically in Fig. 2.

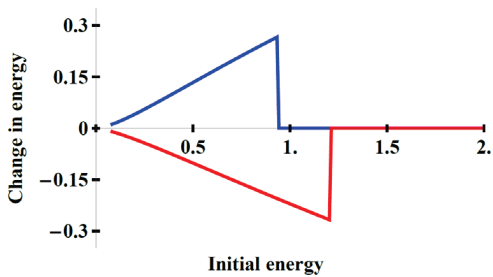

**FIG. 5. Change in the kinetic energy of an atom.** The overall change in the kinetic energy of an atom moving in a focused laser beam due to the dipole force is plotted as a function of its initial energy. The red (lower) curve corresponds to an atom incident from the left while the blue (upper) curve corresponds to an atom incident from the right in analogy with Fig. 2. The initial energy and the change in the energy are both normalized to the kinetic energy of an atom with a velocity of 3.4 km/s in order to simplify the units.

These classical results are in qualitative agreement with the operation of a Maxwell's demon as illustrated in Fig. 2.

## V. NUMERICAL SOLUTION TO SCHRODINGER'S EQUATION

The classical calculations in the previous section suggest that the dipole force could be used to implement a Maxwell's demon. We now compare those results to a fully quantum-mechanical calculation where both the position and the excitation of the atom were treated quantum-mechanically. The standard Hamiltonian for a two-level atom was used and the time-evolution of the wave function was calculated numerically using Mathematica and the Schrodinger equation

$$i\hbar \frac{\partial \psi}{\partial t} = \hat{H}\psi. \qquad (1)$$

The details of the calculation are given in the Appendix.

A typical plot of the real part of the wave function $\psi_G(R,t)$ for the ground-state probability amplitude is shown in Fig. 6 near the point where the velocity of the atom is reversed. It can be seen that the wave function oscillates rapidly as a function of both position and time, which requires a substantial amount of computer memory and execution time. A relatively small value for the mass ($3.3 \times 10^{-31}$ kg) was used in order to keep the de Broglie wavelength of the atom on the order of $1\,\mu$m, which limited the number of oscillations to a manageable value. Although that is smaller than the mass of an actual atom, the force is independent of the mass and this does not affect the qualitative nature of the results.

The results of the quantum-mechanical calculations are summarized in Figs. 4(d) through 4(f), where all of the parameters were the same as in the classical calculations. The initial wave packet was chosen to be a Gaussian with a width (standard deviation) of $5\,\mu$m. The position of the wave packet (as measured by $\langle R \rangle$) is plotted in Fig. 4(d) as a function of time. The absolute value of the velocity of the atom is plotted in Fig. 4(e) as a function of position, where the values along the outgoing trajectory are plotted as red dots since they overlap the blue curve for the incoming trajectory. It can be seen that there is no significant difference between the initial and final magnitude of the velocity or the kinetic energy, unlike the classical results of Figs. 4(b) and 4(c).

These features are consistent with a conservative potential and they show that quantum mechanics does not predict a velocity-dependence optical force on an atom along the axis of a focused laser beam under conditions where incoherent photon scattering is negligible.

## VI. ANALYTIC SOLUTION TO SCHRODINGER'S EQUATION

The classical calculations based on $f_R(R) = -\nabla U(R)$ gave a velocity-dependent force that could be used to implement a Maxwell's demon, while the numerical solution to Schrodinger's equation did not show any velocity dependence. In order to understand the physical origin of this difference, we will now solve for the motion of an atom analytically in the Heisenberg picture. It will be found that the velocity dependence of the dipole force is cancelled out by a force associated with the gradient in the phase of the laser beam.

We will consider a hypothetical single-electron atom with a three-dimensional harmonic oscillator potential instead of a Coulomb potential in order to simplify the calculations. The only significant difference is that the energy levels of a harmonic oscillator are equally spaced, which has no effect if the interaction is sufficiently weak that only the ground state and first excited state have a significant probability of being occupied [17].

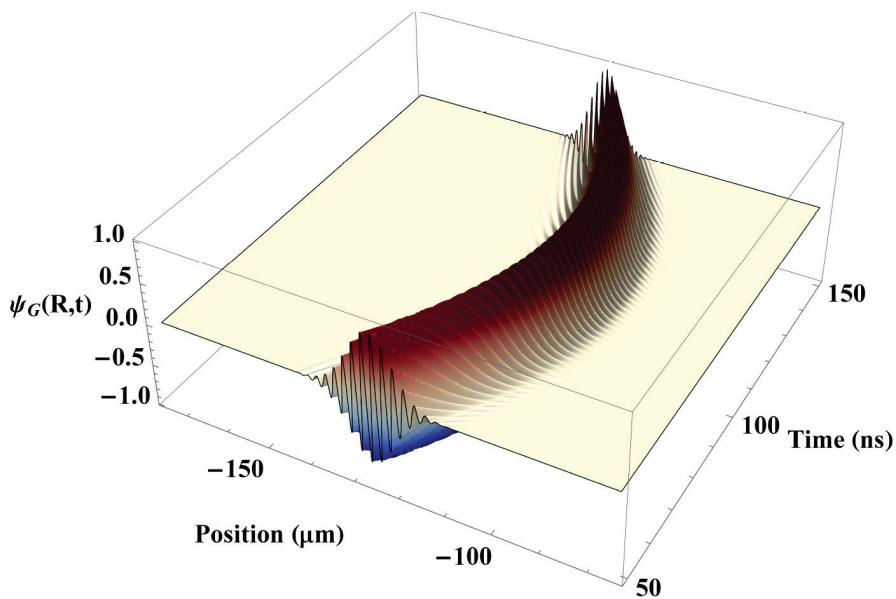

**FIG. 6. Wave function of an atom moving in a focused laser beam.** The real part of the probability amplitude $\psi_G(R,t)$ for an atom to be in its ground state is plotted as a function of position and time, as calculated numerically from Schrodinger's equation. The average position (expectation value) of the atom was calculated from the wave function and used to plot the quantum-mechanical trajectory shown in Fig. 4.

The relative coordinates describing the displacement of the harmonic oscillator will be denoted $x$, $y$, and $z$, with $z$ along the direction of the laser beam, while $R$ will denote the center-of-mass coordinate along the axis of the beam. The dipole moment of the atom is $qy$, where $q$ is the charge of the electron. In the Schrodinger picture, the operators $\hat{y}_S$ and $\hat{R}_S$ are simply the coordinates $y$ and $R$ that appear in the wave function. The corresponding momenta in the Schrodinger picture will be denoted $\hat{p}_S$ and $\hat{P}_S$, respectively.

The Hamiltonian for the atom in the dipole approximation is now

$$\hat{H} = -\frac{\hbar^2}{2m}\frac{\partial^2}{\partial R^2} - \frac{\hbar^2}{2m_e}\nabla_r^2 \\ + \frac{1}{2}k_s\left(x^2 + y^2 + z^2\right) - qy\hat{E}_y(R). \quad (2)$$

Here $k_s$ is a constant that would correspond to the "spring constant" in an ordinary harmonic oscillator and we will use the second-quantized electric field operator $\hat{E}_y(R)$ so that the Hamiltonian is not explicitly time-dependent in the Schrodinger representation. Here $m_e$ is the reduced mass (approximately equal to the mass of the electron) while $m$ is the total mass of the atom.

The motion of an atom can be found most easily by making a unitary transformation to the Heisenberg picture, where as usual [24] we define the operator $\hat{y}_H(R,t)$ by

$$\hat{y}_H(R,t) \equiv e^{i\hat{H}(R)t/\hbar} \hat{y}_S e^{-i\hat{H}(R)t/\hbar}. \tag{3}$$

The Heisenberg operators $\hat{p}_H$, $\hat{R}_H$, and $\hat{P}_H$ are defined in the same way. In the Heisenberg picture, the operator $\hat{y}_H(R,t)$ is a function of both position and time since the Hamiltonian is a function of $R$. This reflects the fact that the induced dipole moment is proportional to the electric field, which varies along the length of the laser beam.

The time dependence of the Heisenberg operators can be found by differentiating Eq. (3). Since the operators $\hat{y}_S$ and $\hat{p}_S$ are not explicitly time dependent, this gives the usual equations for their time evolution:

$$\begin{aligned} \frac{d\hat{y}_H}{dt} &= \frac{1}{i\hbar}[\hat{y}_H, \hat{H}] \\ \frac{d\hat{p}_H}{dt} &= \frac{1}{i\hbar}[\hat{p}_H, \hat{H}]. \end{aligned} \tag{4}$$

The form of the commutators is unaffected by a unitary transformation and they can be calculated using the fact that $\hat{p}_S = (\hbar/i)\partial/\partial y$, which gives

$$\begin{aligned} \left[\hat{y}_H, \hat{H}\right] &= \left[\hat{y}_H, \frac{\hat{p}_H^2}{2m_e}\right] = -\frac{\hbar}{i}\frac{\hat{p}_H}{m_e} \\ \left[\hat{p}_H, \hat{H}\right] &= \left[\frac{\hbar}{i}\frac{\partial}{\partial y_H}, -qy_H\hat{E}_y(\hat{R}_H,t) + \frac{1}{2}k_s y_H^2\right] \\ &= -q\frac{\hbar}{i}\hat{E}_y(\hat{R}_H,t) + \frac{\hbar}{i}k_s y_H. \end{aligned} \tag{5}$$

Inserting these commutators into Eq. (4) gives

$$\begin{aligned} \frac{d\hat{y}_H}{dt} &= \frac{\hat{p}_H}{m_e} \\ \frac{d\hat{p}_H}{dt} &= qE_y(\hat{R}_H,t) - k_s \hat{y}_H. \end{aligned} \tag{6}$$

It should be noted that the unitary transformation gives $\hat{R}_H$ as the argument of the electric field, as can be shown using a Taylor series expansion of $\hat{E}_y$. This reflects the fact that the electric field is evaluated at the location of the moving atom, which is the origin of the Doppler shift. Here we have replaced the second-quantized field operator $\hat{E}_y(\hat{R}_H,t)$ with its classical (expectation) value $E_y(\hat{R}_H,t)$, which is an excellent approximation for a high-intensity laser beam.

The electric field of the laser will be assumed to be a Gaussian beam focused at the origin ($R=0$, $y=0$). We only need the electric field along the central axis of the beam where the atoms are assumed to propagate, which is given in the usual paraxial approximation by [26]

$$E_y(R,t) = \frac{1}{2}E_0 F(R) e^{i[k_\gamma R - g(R) - \omega_L t]}. \tag{7}$$

Here $k_\gamma$ is the wave vector of the light, $E_0$ is a constant, and the envelope function $F(R)$ and the Gouy phase $g(R)$ are given by

$$\begin{aligned} F(R) &= \frac{1}{\sqrt{1+(R/L)^2}} \\ g(R) &= \tan^{-1}\left[\frac{R}{L}\right], \end{aligned} \tag{8}$$

where $L$ is the Rayleigh length. The factor of ½ was included in Eq. (7) so that the total amplitude of the classical field (Eq. (7) plus its complex conjugate) is $E_0$. The slowly-varying Gouy phase has no significant effect here and will be ignored in what follows.

If the force on the atom is sufficiently small, an iterative approach can be used in which, to a first approximation, $\hat{R}_H = R_0 + V_0 t$, where $R_0$ and $V_0$ are the initial position and velocity of the atomic wave packet [16]. This approximation neglects the change in the phase of the field experienced by the atom as a result of its acceleration, which is equivalent to assuming that the Doppler shift is approximately constant over sufficiently small time intervals. With this approximation, the exponentials in Eq. (7) become

$$\begin{aligned} \exp[i(k_\gamma \hat{R}_H - \omega_L t)] &= \exp[i(k_\gamma R_0 + k_\gamma V_0 t - \omega_L t)] \\ &= \exp[i(k_\gamma R_0 - \omega' t)]. \end{aligned} \tag{9}$$

Here we have defined $\omega' \equiv \omega_L - k_\gamma V_0$, which is the Doppler shifted laser frequency.

Inserting Eqs. (7) and (9) into Eq. (6) gives the equation of motion for the electric dipole in the form

$$\frac{d\hat{y}_H}{dt} = \frac{\hat{p}_H}{m_e}$$
$$\frac{d\hat{p}_H}{dt} = \frac{q}{2} E_0 F(R_0 + V_0 t) e^{i(k_\gamma R_0 - \omega' t)} - k_s \hat{y}_H. \quad (10)$$

Eq. (10) corresponds to a coupled set of linear differential equations that are driven by the electric field. This suggests that we look for a solution that is proportional to the electric field of Eq. (7), such as

$$\hat{y}_H = a F(R_0 + V_0 t) e^{i(k_\gamma R_0 - \omega' t)}$$
$$\hat{p}_H = b F(R_0 + V_0 t) e^{i(k_\gamma R_0 - \omega' t)}. \quad (11)$$

Here $a$ and $b$ are unknown coefficients that will be determined by inserting Eq. (11) into Eq. (10) and solving for the required values. We have neglected the initial values of $\hat{y}_H$ and $\hat{p}_H$ before the interaction with the laser beam in Eq. (11), since they have no effect on the induced dipole moment and can be ignored.

Differentiating Eq. (11) with respect to time and inserting the results into Eq. (10) gives

$$\frac{1}{F}\frac{dF}{dR} V_0 a - i\omega' a = \frac{b}{m_e}$$
$$\frac{1}{F}\frac{dF}{dR} V_0 b - i\omega' b = \frac{q}{2} E_0 - k_s a. \quad (12)$$

In the limit of a large Rayleigh length, $F(R)$ will be a slowly-varying function of $R$, in which case $(dF/dR)/F$ will be approximately constant over the width of the wave packet. Eq. (12) can then be rewritten as

$$\frac{b}{m_e} + i\tilde{\omega} a = 0$$
$$k_s a - i\tilde{\omega} b = \frac{q}{2} E_0. \quad (13)$$

Here we have defined the complex parameter $\tilde{\omega}$ by

$$\tilde{\omega} \equiv \omega' + i\frac{1}{F}\frac{dF}{dR} V_0. \quad (14)$$

The solution to Eq. (13) is given by

$$a = \frac{-qE_0}{2m_e} \frac{1}{(\tilde{\omega}^2 - \omega_A^2)}$$
$$b = \frac{i\tilde{\omega} q E_0}{2} \frac{1}{(\tilde{\omega}^2 - \omega_A^2)}. \quad (15)$$

Here we have used the fact that the resonant frequency of the harmonic oscillator is $\omega_A = \sqrt{k_s/m_e}$. If the Doppler shift $\delta\omega_D$ and the term proportional to $V_0$ are much smaller than the detuning $\Delta$, then the coefficient $a$ can be expanded to first order in a Taylor series to give

$$a = \frac{-qE_0}{4\omega_0 m_e} \frac{1}{\Delta}\left(1 - \frac{\delta\omega_D + i\frac{1}{F}\frac{dF}{dR}V_0}{\Delta}\right). \quad (16)$$

Inserting the value of $a$ in Eq. (16) into Eq. (11) gives the corresponding displacement of the electron:

$$\hat{y}_H = \frac{-qE_0}{4\omega_0 m_e} \frac{1}{\Delta}\left(1 - \frac{\delta\omega_D + i\frac{1}{F}\frac{dF}{dR}V_0}{\Delta}\right) \\ \times F(R_0 + V_0 t) e^{i(k_\gamma R_0 - \omega' t)}. \quad (17)$$

Since the classical electric field is real, it must include the complex conjugate of Eq. (7) as well. The complex conjugate of the field induces a displacement equal to the complex conjugate of Eq. (17), and the total displacement is $(y + y^*)$.

The force $\hat{f}_H$ on the atom in the Heisenberg picture can now be evaluated using

$$\langle \hat{f}_H \rangle = \left\langle \frac{d\hat{p}_H}{dt} \right\rangle = \frac{1}{i\hbar}\langle[\hat{p}_H, \hat{H}]\rangle = q\left\langle \hat{y}_H \frac{\partial \hat{E}_y}{\partial R}\right\rangle. \quad (18)$$

Differentiating the electric field in Eq. (7) with respect to $R$ gives

$$\frac{\partial E_y}{\partial R} = \frac{1}{2} E_0 e^{i[k_\gamma R_0 - \omega' t]} F\left(ik_\gamma + \frac{1}{F}\frac{dF}{dR}\right) + c.c., \quad (19)$$

where we have included the complex conjugate of the field.

The product of $(y + y^*)$ and $\partial E_y / \partial R$ contains four terms, two of which oscillate at a frequency of $\pm 2\omega'$ and average to zero. The remaining two terms are given by

$$\langle \hat{f}_H \rangle = \frac{-q^2 E_0^2}{8\omega_0 m_e} \frac{F^2}{\Delta} \left( 1 - \frac{\delta\omega_D - i\frac{1}{F}\frac{dF}{dR}V_0}{\Delta} \right)$$

$$\times \left( ik_\gamma + \frac{1}{F}\frac{\partial F}{\partial R} \right) + c.c. \quad (20)$$

Multiplying out the terms in Eq. (20) gives

$$\langle \hat{f}_H \rangle = \frac{-q^2 E_0^2}{4\omega_0 m_e} \frac{F}{\Delta} \frac{dF}{dR} \left( 1 - \frac{(\delta\omega_D + k_\gamma V_0)}{\Delta} \right). \quad (21)$$

Given the fact that $\delta\omega_D = -k_\gamma V_0$, it can be seen that the last two terms in Eq. (21) cancel out and the force on the atom reduces to

$$\langle \hat{f}_H \rangle = \frac{-q^2 E_0^2}{4\omega_0 m_e} \frac{F}{\Delta} \frac{dF}{dR}. \quad (22)$$

The total force in Eq. (22) only depends on the detuning $\Delta = \omega_\gamma - \omega_0$ for an atom at rest, and the net force is independent of the velocity. The velocity dependence of the dipole force due to the $\partial F / \partial R$ term in Eq. (19) is cancelled out by the force due to the $ik_\gamma$ term, at least to lowest order. The $\partial F / \partial R$ term corresponds to the usual dipole force due to the gradient in the intensity, while the $ik_\gamma$ term comes from the gradient in the phase of the laser beam.

## VII. DISCUSSION AND CONCLUSIONS

These results show that a Maxwell's demon cannot be implemented using a focused laser beam and the Doppler shift as suggested in Fig. 2. If that were not the case, it would be possible to implement a Maxwell's demon that dissipated a negligible amount of energy.

The analytic calculations in the previous section show that the velocity-dependent contributions from the gradient in the intensity and the gradient in the phase of the laser beam cancel out, giving a total force along the direction of the laser beam that is independent of velocity. This cancellation is dependent on the imaginary term in Eq. (16) for the coefficient $a$, which is proportional to the induced dipole moment. There is no net energy transfer between the field and a stationary atom (assuming that the incoherent scattering is negligible), which would require that the electric field and the current be $90^0$ out of phase. But the motion of the atom causes a change in the population of its excited state with a corresponding absorption of a virtual photon. This requires that the electric field and the current not be entirely $90^0$ out of phase, as reflected by the imaginary term in Eq. (16). Without this phase lag, the imaginary $ik_\gamma$ term in the expression for the force in Eq. (20) would give no contribution after adding the complex conjugate.

From a quantum-mechanical point of view, the $ik_\gamma$ contribution to the force can be interpreted as the recoil momentum from the absorption of a single virtual photon; that momentum is returned to the field when the atom leaves the focal region provided that the process is adiabatic. Classically, this can be understood as the magnetic force produced by the current in an atom interacting with the magnetic field of the laser beam. Once again, the current and the magnetic field would be $90^0$ out of phase without the imaginary term in Eq. (16), giving zero net force from the $ik_\gamma$ term.

The contribution to the force due to the $ik_\gamma$ term has been discussed previously [15-17, 20] and it is sometimes referred to as the dissipative force because of its velocity dependence [15, 16]. Here we have shown that the velocity dependence of the dipole force is cancelled out to lowest order by the contribution from the $ik_\gamma$ term under conditions where photon scattering is negligible. This is necessary to avoid the possibility of constructing a Maxwell's demon that dissipates negligible energy.

So far we have only considered the force along the direction of the laser beam. There will be a dipole force perpendicular to the direction of propagation if the atom is located off the central axis of the beam, where there is a gradient of the intensity in the transverse direction. There need not be any gradient of the phase in that direction, in which case the $ik_\gamma$ force no longer cancels the velocity dependence of the dipole force. In fact, velocity-dependent dipole forces have been experimentally observed, but only in the direction transverse to the axis of the laser beam [12-14]. The force in the transverse direction cannot be used to implement a Maxwell's demon as illustrated in Fig. 2, however.

The calculations in the previous section only included the lowest-order term in the Taylor series expansion of Eq. (16). A more detailed calculation shows that the velocity dependences of these two forces do not cancel out for higher-order terms, with a net velocity dependence that is proportional to $1/\Delta^3$ along the axis of the laser beam. Although this would allow the operation of a relatively inefficient Maxwell's demon, incoherent scattering and non-adiabatic effects would no longer be negligible in comparison.

In summary, the existence of a velocity-dependent optical force with negligible dissipation would allow the implementation of a Maxwell's demon capable of reducing the overall entropy. We have shown that the velocity dependence of the dipole force along the direction of propagation of a focused laser beam is cancelled out by another force related to the gradient in the phase of the field rather than the gradient in its intensity. These results provide a further illustration of the fact that the operation of a Maxwell's demon requires the dissipation of a minimum amount of energy on the order of $kT$.

## ACKNOWLEDGEMENTS

This work was supported in part by the National Science Foundation under grant # PHY-1802472.

## APPENDIX A: NUMERICAL CALCULATIONS

This Appendix provides a more detailed description of the numerical calculations, including the classical trajectory and the solution to Schrodinger's equation.

The classical calculations were based on Newton's laws and the dipole force $f(R) = -\nabla U(R)$, where $U(R)$ is the effective potential energy due to the dipole interaction. $U(R)$ was calculated using lowest-order perturbation theory as described in the text. The Doppler shift in the frequency of the laser beam as seen by the atom was included in the calculation of $U(R)$. The corresponding set of differential equations is

$$\frac{dR}{dt} = \frac{P}{m}$$
$$\frac{dP}{dt} = f(R), \quad (A1)$$

where $P$ is the classical momentum of the atom.

The electric field of the laser was assumed to be given by the Gaussian beam described in Eq. (7) in the text. The coupled differential equations in Eq. (A1) were solved numerically using the NDSolve routine in Mathematica. Results with 6 significant digits of precision could be readily obtained since the force only depends on $|M|^2$ and does not contain any rapidly-oscillating terms. The choice of parameters used in the plots shown in Fig. 4(a) through 4(c) are given in the text and they are the same parameters used in the quantum-mechanical calculations discussed below.

The quantum-mechanical results shown in Figs. 4(d) through 4(f) were obtained by numerically integrating Schrodinger's equation. Using center-of-mass coordinate $R$ and relative coordinates $\mathbf{r}$ allows Schrodinger's equation to be written as

$$i\hbar \frac{\partial \psi(R,\mathbf{r},\mathbf{t})}{\partial t} = \hat{H}\psi = -\frac{\hbar^2}{2m}\frac{\partial^2 \psi}{\partial R^2} - \frac{\hbar^2}{2m_e}\nabla^2_\mathbf{r}\psi + V(\mathbf{r}) - qyE_y(R). \quad (A2)$$

Here $m_e$ is the reduced mass (very nearly equal to the electron mass) and $V(\mathbf{r})$ is the Coulomb potential, where we have assumed a hydrogen-like atom with a single electron for simplicity. We can simplify the calculations by expanding the wave function in the unperturbed eigenfunctions of the relative-coordinate part of the Hamiltonian (the usual atomic states). If the laser beam is close to resonance with only a single excited state, this allows Schrodinger's equation to be rewritten in the form [18]

$$i\hbar \frac{\partial \psi_G(R,t)}{\partial t} = -\frac{\hbar^2}{2m}\frac{\partial^2 \psi_G}{\partial R^2} + E_G\psi_G + M^*\psi_E$$
$$i\hbar \frac{\partial \psi_E(R,t)}{\partial t} = -\frac{\hbar^2}{2m}\frac{\partial^2 \psi_E}{\partial R^2} + E_E\psi_E + M\psi_G. \quad (A3)$$

Here $\psi_G(R,t)$ and $\psi_E(R,t)$ are the probability amplitudes for an atom to be in the ground or excited states. The matrix element $M = \langle \phi_E | -qyE_y(R,t) | \phi_G \rangle$ as usual, and this includes the $\exp[i(kR - g(R) - \omega_L t)]$ factor in the electric field. The recoil energy of the atom is automatically included in the kinetic energy terms in Eq. (A3).

The probability amplitudes $\psi_G(R,t)$ and $\psi_E(R,t)$ both oscillate at optical frequencies due to the $E_G$ and $E_E$ terms in Eq. (A3). Much of this oscillation can be removed by defining $\psi'_G = \exp[iE_G t/\hbar]\psi_G$, $\psi'_E = \exp[iE_E t/\hbar]\psi_E$, and $E'(R,t) = \exp[i(E_E - E_G)/\hbar]E(R,t)$ (the interaction picture). The new amplitudes $\psi'_G(R,t)$ and $\psi'_E(R,t)$ satisfy Eq. (A3) with $E_G = E_E = 0$ and the field oscillating at an angular frequency of $\Delta$.

This coupled set of differential equations for $\psi'_G$ and $\psi'_E$ was also solved numerically using the Mathematica NDSolve routine. Unlike the classical equations, the matrix element $M$ that appears in Eq. (A3) is a rapidly oscillating function of position and time and the probability amplitudes $\psi'_G(R,t)$ and $\psi'_E(R,t)$ are also rapidly oscillating as a result. The wavelength of the laser beam corresponded to a characteristic length of 1 $\mu$m and a detuning of 5 GHz corresponded to a characteristic time scale of approximately 0.2 ns. The grid spacings in the NDSolve routine were necessarily smaller than these characteristic length and time scales. In order for the process to be adiabatic, the Rayleigh length was chosen to be 100 $\mu$m and the time scale for the motion of an atom through the laser beam was on the order of 300 ns. Thus a large number of grid points was required and the computer program required up to 40 Gigabytes of random access memory (RAM).

The numerical calculations were performed on a personal computer/work station with 96 Gigabytes of RAM and 8 processor threads (two processor chips). The calculations typically required an hour of execution time. Mathematica stored the results for all values of $R$ and $t$ in a large interpolation array. The memory requirements could be reduced by writing custom software that only stored the results at periodic time intervals but that was unnecessary.

The precision of the numerical results was estimated by varying the parameters in the NDSolve routine, such as the minimum grid spacing, and comparing the results of the calculations. The precision was also checked by comparing the norm of the wave function at the beginning and end of the calculations. It was estimated that the results were accurate to 4 or 5 significant digits.

The calculated rate of oscillation in time of the real part of $\psi'_G(R,t)$ was actually an order of magnitude faster than the rate of oscillation shown in Fig. 6. In order to plot the oscillations with enough resolution for them to be visible, it was necessary to multiply $\psi'_G(R,t)$ by a factor of $\exp[i\Omega t]$ where the constant $\Omega$ was chosen to cancel out most of the oscillation. This has no effect on the results of the calculations and this factor was not applied to the data shown in Fig. 4. The calculations were performed for several different sets of parameters, and the parameters used in Fig. 6 are slightly different from those used to obtain the data points shown in Fig. 4.

The mass of the atom was arbitrarily taken to be $3.31 \times 10^{-31}$ kg so that the de Broglie wavelength of an atom was on the order of 1 $\mu$m. This was necessary to limit the grid size to a manageable value. Using a more realistic value of the mass would have increased the grid size by approximately four orders of magnitude and the calculations would not have been feasible. The dipole force on an atom is independent of its mass and only the acceleration is mass dependent. The choice of the mass was not important here since the main purpose of these calculations was to determine whether or not the dipole force is velocity dependent.

The results of Figs. 4(e) and 4(f) show that the decrease in the final velocity and kinetic energy are much smaller than is the case in the classical calculations based on the dipole force alone, and we can conclude that quantum mechanics does not predict a velocity-dependent force along the direction of propagation of the beam. The velocity data shown in Fig. 4(e) was obtained by numerical differentiation of the calculated position. Although the data in the figure for the outgoing portion of the trajectory appear to lie on top of the incoming trajectory, the final velocity was actually 0.4% less than the initial velocity. This can be attributed to the fact that the adiabatic theorem was only approximately satisfied for the parameters used here. As a result, there is a small probability that an atom will be left in the excited state at the end of the process, with a net loss of one photon. This effect can be made arbitrarily small by increasing the Rayleigh length, but that would have increased the grid size and the memory requirements by an unacceptable amount.